\documentstyle[12pt]{article}

\newcommand{\br}{{\bf r}}
\newcommand{\be}{\begin{equation}}
\newcommand{\ee}{\end{equation}}

\newcommand{\bt}{\beta}
\newcommand{\al}{\alpha}

\newcommand{\ra}{\rightarrow}

\begin{document}

\begin{center}
{\large{\bf Coherence phenomena}} \\ [5mm]
{\large{\bf V.I. Yukalov$^{1,2}$}} \\ [3mm]
{\it $^1$Bogolubov Laboratory of Theoretical Physics, \\
Joint Institute for Nuclear Research, Dubna 141980, Russia\\ [2mm]
$^2$Institut f\"ur Theoretische Physik, \\
Freie Universit\"at Berlin, Arnimallee 14, D-14195 Berlin, Germany}
\end{center}

\vskip 10mm

The word coherence comes from the Latin {\it cohaerens} which
means {\it being in relation}. This implies that several objects
are in some sense interrelated or correlated with each other.
Coherence phenomena are those displaying a high level of
correlation between several objects.

From the physical point of view, it is necessary to distinguish between
two types of coherence, which may be named state coherence and transition
coherence. The {\it state coherence} characterizes correlations between
static properties of the considered objects, while the {\it transition
coherence} describes correlated dynamical processes. When mentioning
coherence, one very often keeps in mind solely the second type, that is,
the transition coherence which is related to radiation processes. In so
doing, one forgets about the state coherence. Or, in the best case, these
two types of coherence are treated separately, as being divorced from
each other. However, both these types of coherence are two sides of the
same story, and one can infer a more correct insight from considering
them together.

To catch an intuitive idea of these two types of coherence, one
can imagine the following picture. Let a group of soldiers stay
being immovable, all of them keeping the same position. This is
what corresponds to state coherence. If all soldiers would keep
different positions, some staying, some sitting, some laying, there
would be no state coherence between them. Now, if one conceives
well aligned rows of soldiers in a parade, moving synchronously
with respect to each other, one should say that they are
displaying transition coherence. And if they would march with
different velocities and in different directions, transition
coherence would be absent.

Coherence, being the property of well correlated objects, is,
clearly, also related to the existence of a kind of order. Be it a
static order defining the same positions or an ordered motion of a
group. Then the notion, opposite to coherence, should also be an
antonym to order, that is, chaos. Thus, the {\it state chaos}
means the absence of any static order among several objects, and
the {\it transition chaos} implies an absolutely disorganized
motion of an ensemble constituents.

The notion of coherence is implicit in the existence of
correlation among several objects. The latter can be enumerated
with the index $i=1,2,\ldots,N$. Each object, placed in the
spatial point ${\bf r}_i$, at time $t$, can be associated with a
set $\{ Q_\alpha({\bf r}_i,t\}$ of observable quantities labelled
by $\alpha$. To formalize the definition of the state and
transition coherence, one may proceed as follows. For brevity, one
may write $Q_i^\al=Q_\al({\bf r}_i,t)$. Let $Q_i^z$ correspond to a
state property of an object, while $Q_i^x$ and $Q_i^y$ describe
its motion. As an illustration, one could keep in mind that
$Q_i^\al$ are the spin components. Another example could be in
assuming that $Q_i^z$ is the population difference of a resonant
atom, while $Q_i^x$ and $Q_i^y$ are its transition dipoles. Instead of
considering the latter separately, it is convenient to introduce
the complex combinations $Q_i^\pm\equiv Q_i^x\pm iQ_i^y$. In
general, $Q_i^\al$ are not simply classical quantities but are
operators. If the considered system is associated with a
statistical operator $\hat\rho$, then the observable quantities
are the {\it statistical averages}
\be
\label{1}
<Q_i^\al>\; \equiv {\rm Tr}\hat\rho \; Q_i^\al \; ,
\ee
expressed by means of the trace operation. A handy way of describing the
system features is by introducing the dimensionless quantities, normalized
to the number of objects $N$ and to the maximal value $Q\equiv\max<Q_i^z>$.
Then one may define the {\it state variable}
\be
\label{2}
s\equiv \frac{1}{QN} \sum_{i=1}^N <Q_i^z>
\ee
and the {\it transition variable}
\be
\label{3}
 u \equiv \frac{1}{QN} \sum_{i=1}^N <Q_i^-> \; .
\ee
In the possible collective state of a system, one may
distinguish two opposite cases, when the individual states of all
objects are the same and when they are randomly distributed. These
two limiting cases give
\begin{eqnarray}
\label{4}
|s| =\left\{ \begin{array}{cc}
1, & state \; coherence \\
\\
0, & state \; chaos \; . \end{array}
\right.
\end{eqnarray}
Considering the transition characteristic (3), one may keep in
mind collective motion of an ensemble of oscillators. Then again
there can be two opposite situations, when the oscillation
frequencies of all oscillators, as well as their initial phases,
are identical and when these take randomly different values. For
the corresponding limiting cases of completely synchronized
oscillations and of an absolutely random motion, respectively, one
has
\begin{eqnarray}
\label{5}
|u| =\left\{ \begin{array}{cc}
1, & transition \; coherence \\
\\
0, & transition \; chaos \; . \end{array}
\right.
\end{eqnarray}
In the intermediate situation, one may say that there is {\it
partial state coherence} if $0<|s|<1$ and that there occurs {\it
partial transition coherence} when $0<|u|<1$.

Accepting that coherence is not compulsory total, when all parts
of a system are perfectly correlated, but that it can be partial,
one comes to the necessity of defining qualitative characteristics
for such a partial coherence. For this purpose, since coherence
and correlation are intimately interrelated, one introduces
correlation functions. Let $Q_\al^+(\br,t)$ denote Hermitian
conjugation for an operator $Q_\al(\br,t)$. When $Q_\al(\br,t)$ is
a nonoperator function, Hermitian conjugation means complex
conjugation. For any two operators from the set $\{
Q_\al(\br,t)\}$ one may define the {\it correlation function}
\be
\label{6} C_{\al\bt}(\br_1,t_1,\br_2,t_2) \equiv \;
<Q_\al^+(\br_1,t_1)Q_\bt(\br_2,t_2)> \; . \ee The function
$C_{\al\al}(\ldots)$ for coinciding operators is called {\it
autocorrelation function}. One also uses the shifted correlation
function $$ B_{\al\bt}\equiv \; <Q_\al^+Q_\bt>-<Q_\al^+><Q_\bt>
\;, $$ where, for brevity, the spatio-temporal variables are not
written down explicitly. For describing coherent processes, one
often employs the normalized correlation function
\be
\label{7}
 K_{\al\bt} \equiv \frac{<Q_\al^+ Q_\bt>}{(<Q_\al^+
Q_\al><Q_\bt^+ Q_\bt>)^{1/2}} \; , \ee which sometimes is termed
{\it coherence function}. The functions (6) and (7) can be
specified as {\it second-order correlation functions} since, in
general, it is possible to define higher-order correlation
functions, such as the $2p$-order function $$
C_{\al_1\ldots\al_{2p}}= \; <Q_{\al_1}^+\ldots Q_{\al_p}^+
Q_{\al_{p+1}}\ldots Q_{\al_{2p}}> \; . $$ Such correlation
functions are closely related to reduced density matrices.

Correlations are usually strongest at the closest spatio-temporal points.
Thus, function (7) varies in the interval $0\leq|K_{\al\bt}|\leq 1$,
being maximal for the autocorrelation function $|K_{\al\al}|=1$ at the
coinciding points $\br_1=\br_2$, $t_1=t_2$. When either spatial or
temporal distance between two points increases, correlations diminish,
which is named {\it correlation decay}. At asymptotically large distance,
the correlation function (6) for two local observables displays the
property of correlation weakening or {\it correlation decoupling}
\be
\label{8}
<Q_\al^+(\br_1,t_1)Q_\bt(\br_2,t_2)>\; \simeq \;
<Q_\al^+(\br_1,t_1)><Q_\bt(\br_2,t_2)> \; ,
\ee
where either $|\br_1-\br_2|\ra\infty$ or $|t_1-t_2|\ra\infty$. It is
important to stress that property (8) holds only for local observables.
But for operators representing no observable quantities, such a property
of correlation decoupling generally has no sense.

Referring to coherence, one characteristically implies correlations
between similar objects, which requires the usage of autocorrelations
functions. Describing coherence decay also needs to fix a point from
which this decay is measured. It is customary to place the reference
point at $\br=0$ and $t=0$ and to study coherence decay by considering an
autocorrelation function
\be
\label{9}
C_\al(\br,t) \equiv\; <Q_\al^+(\br,t)\; Q_\al(0,0)> \; .
\ee
In many cases, there exists a spatial direction of particular importance.
This, e.g., could be the direction of field propagation. Then it is natural
to associate this special direction with the longitudinal $z$-axis and
the transverse direction with the radial variable $r_\perp$. The
characteristic scale of coherence decay in the longitudinal direction is
called {\it coherence length} $l_{coh}$,
\be
\label{10}
l_{coh}^2 \equiv \frac{\int z^2|C_\al(\br,t)|^2\;d\br}
{\int |C_\al(\br,t)|^2\; d\br} \; ,
\ee
where the integration is over the whole space volume. Coherence decay in
the transverse direction is classified as {\it transverse coherence radius}
$r_{coh}$
\be
\label{11}
r_{coh}^2 \equiv \frac{\int
r_\perp^2|C_\al(\br,t)|^2\;d\br} {\int|C_\al(\br,t)|^2\; d\br}\; .
\ee
For isotropic systems, one replaces $r_\perp$ by the spherical radius $r$
and obtains from equation (11) {\it coherence radius}. It is straightforward
to call $A_{coh}\equiv\pi r_{coh}^2$ {\it coherence area} and
$V_{coh}\equiv A_{coh}l_{coh}$, {\it coherence volume}. The typical scale
of temporal correlation decay is termed {\it coherence time} $t_{coh}$,
\be
\label{12}
t_{coh}^2 \equiv \frac{\int_0^\infty
t^2|C_\al(\br,t)|^2\;dt} {\int_0^\infty|C_\al(\br,t)|^2\; dt}\; .
\ee
As is seen, the coherence length (10) and coherence radius (11) are related
to a fixed moment of time, while the coherence time (12) defines the temporal
coherence decay at a given spatial point. All equations (10)-(12) have to do
with a particular coherence phenomenon characterized by the correlation
function (9).

{\it Phase transitions} in equilibrium statistical systems are the collective
phenomena demonstrating a variety of different types of {\it state coherence}
arising under adiabatically slow variation of thermodynamic or system
parameters. The latter can be temperature, pressure, external fields, and
so on. A phase transition is a transformation between different thermodynamic
phases that are conventionally specified by means of {\it order parameters},
which are defined as statistical averages of operators corresponding to some
local observables. The order parameter is assumed to be zero in a disordered
phase, while nonzero in an ordered phase. For example, the order parameter
at Bose-Einstein condensation is the fraction or density of particles in the
single-particle ground state. The order parameter for superconducting phase
transition is the density of Cooper pairs or the related gap in the excitation
spectrum. Superfluidity is characterized by the fraction or density of the
superfluid component. For magnetic phase transitions, the order parameter
is the average magnetization. Thermodynamic phases can also be classified by
{\it order indices}. Let the autocorrelation function (9) be defined for the
operator related to an order parameter. Then, for a disordered phase, the
coherence length is close to the interparticle distance and the coherence
time is about the interaction time. But for an ordered phase, the coherence
length is comparable with the system size and the coherence time becomes
infinite.

In the quasiequilibrium picture of phase transitions, taking account of
heterophase fluctuations, there appear {\it mesoscopic coherent structures},
with the coherence length much larger than interparticle distance but much
smaller than the system size. The coherence time of these mesoscopic coherent
structures, which is their lifetime, is much longer than the local equilibrium
time, though may be shorter than the observation time. Such coherent
structures are similar to those arising in turbulence.

Electromagnetic {\it coherent radiation} by lasers and masers presents
a perfect example of {\it transition coherence}. Such radiation processes
are accompanied by {\it interference patterns}. Interference is a phenomenon
typical of coherent radiation. The latter can be produced by atoms, molecules,
nuclei or other radiating objects. Interference effects caused by light
beams are studied in nonlinear optics. But coherent radiation, and related
interference effects, also exist in other diapasons of electromagnetic
radiation frequencies, e.g., in infrared-, radio-, or gamma-regions. Moreover,
there exist other types of field radiation, such as acoustic radiation or
emission of matter waves formed by Bose-condensed atoms. Registration of
interference between a falling beam and that reflected by an object is the
basis for holography that is the method or recording and reproducing wave
fields. The description of interference involves the usage of correlation
functions. Let $Q_i(t)$ represent a field at time $t$, produced by a radiator
at a spatial point $\br_i$. The radiation intensity of a single emitter may
be defined as
\be
\label{13}
I_i(t) \equiv \; <Q_i^+(t)\; Q_i(t)> \; .
\ee
Then the radiation intensity for an ensemble of $N$ emitters writes
\be
\label{14}
I(t)=\sum_{i,j=1}^N <Q_i^+(t)\; Q_j(t)> \; .
\ee
Separating here the sums with $i=j$ and with $i\neq j$ results in
\be
\label{15}
I(t) = \sum_{i=1}^N I_i(t) + \sum_{i\neq j}^N <Q_i^+(t)\; Q_j(t)> \; ,
\ee
which shows that intensity (14) is not simply a sum of the intensities (13) of
individual emitters, but also includes the interference part, expressed through
the autocorrelation functions of type (9). The first term in equation (15) is
the intensity of incoherent radiation, while the second term corresponds to the
intensity of coherent radiation.

Coherence phenomena, related both to state coherence and transition coherence,
have found wide use in various applications.

\vskip 10mm

{\it See also} Bose-Einstein condensation; Chaotic dynamics;
Critical phenomena; Dynamical systems; Feedback; Ferroelectricity
and ferromagnetism; Lasers; Maxwell-Bloch system; Nonequilibrium
statistical mechanics; Nonlinear acoustics; Nonlinear optics;
Order parameters; Pattern formation; Phase transitions; Quantum
chaos; Quantum nonlinearity; Spatio-temporal chaos; Spin systems;
Structural complexity; Superconductivity; Superfluidity;
Turbulence

\vskip 5mm

{\bf Further Reading}

\vskip 2mm

Andreev, A.V., Emelyanov, V.I., and Ilinski, Y.A. 1993. {\it
Cooperative Effects in Optics}, Bristol: Institute of Physics

\vskip 2mm

Benedict, M.G., Ermolaev, A.M., Malyshev, V.A., Sokolov, I.V. and
Trifonov, E.D. 1996. {\it Superradiance: Multiatomic Coherent
Emission}, Bristol: Institute of Physics

\vskip 2mm

Bogolubov, N.N. 1967. {\it Lectures on Quantum Statistics}, Vol.
1, New York: Gordon and Breach.

\vskip 2mm

Bogolubov, N.N. 1970. {\it Lectures on Quantum Statistics}, Vol.
2, New York: Gordon and Breach.

\vskip 2mm

Coleman, A.J. and Yukalov, V.I. 2000. {\it Reduced Density
Matrices}, Berlin: Springer

\vskip 2mm

Klauder, J.R. and Skagerstam, B.S. 1985. {\it Coherent States},
Singapore: World Scientific

\vskip 2mm

Klauder, J.R. and Sudarshan, E.C.G. 1968. {\it Fundamentals of
Quantum Optics}, New York: Benjamin

\vskip 2mm

Lifshitz, E.M. and Pitaevskii, L.P. 1980. {\it Statistical
Physics: Theory of Condensed State}, Oxford: Pergamon

\vskip 2mm

Mandel, L. and Wolf, E. 1995. {\it Optical Coherence and Quantum
Optics}, Cambridge: Cambridge University

\vskip 2mm

Nozi\`eres, P. and Pines, D. 1990. {\it Theory of Quantum Liquids:
Superfluid Bose Liquids}, Redwood: Addison-Wesley

\vskip 2mm

Perina, J. 1985. {\it Coherence of Light}, Dordrecht: Reidel

\vskip 2mm

Scott, A.C. 1999. {\it Nonlinear Science: Emergence and Dynamics
of Coherent Structures}, Oxford: Oxford Univerisity

\vskip 2mm

Ter Haar, D. 1977. {\it Lectures on Selected Topics in Statistical
Mechanics}, Oxford: Pergamon

\vskip 2mm

Yukalov, V.I. 1991. Phase transitions anad heterophase
fluctuations. {\it Physics Reports}, 208: 395--492

\vskip 2mm

Yukalov, V.I. and Yukalova, E.P. 2000. Cooperative electromagnetic
effects. {\it Physics of Particles and Nuclei}, 31: 561--602

\end{document}